\documentclass{moriond}

\def\be{\begin{equation}}
\def\ee{\end{equation}}
\def\bea{\begin{eqnarray}}
\def\eea{\end{eqnarray}}

\newcommand{\lsim}{\stackrel{<}{_\sim}}

\newcommand{\Photo}{}

\begin{document}
\vspace*{4cm}
\title{LEPTON NONUNIVERSALITY  ANOMALIES \&  IMPLICATIONS}

\author{ G.HILLER }

\address{Fakult\"at  Physik, TU Dortmund, Otto-Hahn-Str.4, D-44221 Dortmund, Germany}

DO-TH 18/07

\maketitle\abstracts{
We discuss avenues for  diagnosing new physics hinted from lepton nonuniversality in rare $b$-decays, and physics implications. }

\section{The situation}

We are presently seeing $\sim 2.6 \sigma$  
hints of new physics (NP)  in rare semileptonic $b \to s ll$ transitions,  indicating lepton nonuniversality (LNU)  between electrons and muons
in  each observable $R_K$ and $R_{K^*}$   \cite{Hiller:2003js,Aaij:2014ora,Aaij:2017vbb} \cite{albrecht}
\begin{equation}
 R_{H} =\frac{ \int_{\rm q^2_{\rm min}}^{q^2_{\rm max}} d q^2 \, d{\cal{B}}/dq^2 (\bar B \to \bar H \mu \mu) }{ \int_{\rm q^2_{\rm min}}^{q^2_{\rm max}} d q^2  \, d {\cal{B}}/dq^2  (\bar B \to \bar H e e) }  \, ,  ~~ H=K,K^*,X_s,...
\end{equation}
\begin{eqnarray}
R^{\rm{LHCb}}_K&=0.745^{+0.090}_{-0.074}\pm 0.036 \, , ~~~ R^{\rm{LHCb}}_{K^\ast}=0.69^{+0.11}_{-0.07} \pm 0.05    \label{RKLHCb}
\end{eqnarray}
for  the dilepton mass cuts  $q^2_{\rm min}= 1.1 \,\rm{GeV}^2$ for $R_{K^\ast}$ and  $ 1 \,\rm{GeV}^2$ for $R_{K}$, and $q^2_{\rm max}= 6 \,\rm{GeV}^2$.
In lepton-universal models  including the SM holds $R_H=1$ up to tiny  corrections of ${\cal{O}}(m_\mu^2/m_b^2)$ despite of the sizable hadronic uncertainties  in the individual rates \cite{Hiller:2003js}. Electromagnetic  corrections \cite{Huber:2005ig} \cite{Bobeth:2007dw} are found to not exceed percent level \cite{Bordone:2016gaq}.
$R_H-1$ are clean null tests of the standard model (SM). 
Previous measurements of  $R_{K,K^*}$   by Belle and BaBar  are consistent with one.
We discuss 
in section \ref{sec:MIA}  which operators can be responsible for the deviation (\ref{RKLHCb})  from universality \cite{Hiller:2014yaa} \cite{Hiller:2014ula}.
In section \ref{sec:evsmu} lepton-specific measurements are emphazised as a means  to understand whether the present LNU anomalies are due to 
physics beyond the standard model (BSM) in electrons, in muons, or in both \cite{fleischer,crocombe}, and CP violation is commented on.
We discuss  side effects from flavor \cite{Glashow:2014iga} \cite{Varzielas:2015iva} in section \ref{sec:side}, which addresses correlations with other sectors, such as charm, or Kaon physics \cite{iyer,GD}, as well as lepton flavor violation (LFV),  and decays with $\tau$'s, or $\nu$'s.
Collider implications and leptoquark signatures related to the $b$-decay anomalies are discussed in section \ref{sec:LQ}  \cite{you,morse,crivellin}.
We comment on the status of $R_{D,D^*}$ in section \ref{sec:RD}.

\section{Model-independent analysis \label{sec:MIA}}

One employs an effective low energy theory $
{\cal{H}}_{\rm eff}= -4 \frac{G_F}{ \sqrt{2}}  \, V_{tb} V_{ts}^* \, 
\sum_i C_i(\mu) O_i(\mu)  $ at dimension six 
\begin{eqnarray} \mbox{V,A operators}
~~  {\cal{O}}_{9} & = & [\bar{s} \gamma_\mu P_{L} b] \, [\bar{\ell} \gamma^\mu \ell] \,, \quad  {\cal{O}}_{9}^\prime  =  [\bar{s} \gamma_\mu P_{R} b] \, [\bar{\ell} \gamma^\mu \ell]  \, ,  \\
~~  {\cal{O}}_{10} & = &[\bar{s} \gamma_\mu P_{L} b] \, [\bar{\ell} \gamma^\mu \gamma_5 \ell] \,,  \quad  {\cal{O}}_{10}^\prime  = [\bar{s} \gamma_\mu P_{R} b] \, [\bar{\ell} \gamma^\mu \gamma_5 \ell] \, ,  \\
\mbox{S,P operators}
~~  {\cal{O}}_{S} & = & [\bar{s}  P_{R} b] \, [\bar{\ell} \ell] \,, \quad   {\cal{O}}_{S} ^\prime =  [\bar{s}  P_{L} b] \, [\bar{\ell} \ell] \,,\\
~~ {\cal{O}}_{P} & =&  [\bar{s}  P_{R} b] \, [\bar{\ell}  \gamma_5 \ell ]\,,  \quad {\cal{O}}_{P}^\prime  = [\bar{s}  P_{L} b] \, [\bar{\ell}  \gamma_5 \ell ]  \, , \\
\mbox{ tensors} 
~~   {\cal{O}}_T &  = & [\bar{s} \sigma_{\mu\nu} b] \, [\bar{\ell} \sigma^{\mu\nu} \ell]\,, \quad
   {\cal{O}}_{T5}  = 
%\frac{i}{2}\, \varepsilon^{\mu\nu\alpha\beta}
    [\bar{s} \sigma_{\mu\nu} b]\, [\bar{\ell} \sigma^{\mu\nu}\gamma_5 \ell]  \, . 
    \end{eqnarray}
    This set of semileptonic operators is complete. To discuss LNU one needs to add lepton specific indices $C_i O_i  \rightarrow C_i^\ell O_i^\ell  $, $\ell=e, \mu,\tau$.
    In the SM, only $O_9,O_{10}$  receive non-negligible and universal contributions, $C_9^{\rm SM} \simeq -C_{10}^{\rm SM} \simeq 4.1$, all other operators are BSM-induced.
        
To interpret LNU data (\ref{RKLHCb})  it is useful to employ the approximation where BSM physics enters the branching ratios linearly,
schematically, with amplitude $A=A^{\rm SM}+A^{\rm NP}$,
\begin{eqnarray}
 {\cal{B}}=|A|^2=| A^{\rm SM}|^2 + 2  \, {\rm Re}( A^{\rm SM} A^{{\rm NP}^*}) + | A^{\rm NP}|^2    \, ,    \label{eq:B}
\end{eqnarray}
 that is, assuming $|C^{\rm NP}| \ll  |C^{\rm SM}|$.
The complementarity between $R_K$ and $R_{K^*}$ becomes manifest \cite{Hiller:2014ula}. In fact, 
 it suffices to measure two  different (by spin parity of the final hadron) $R_H$ ratios. Then, all others
serve as consistency checks, because the  Wilson coefficients $C$ and $C^\prime$
enter  decay amplitudes in specific combinations dictated by parity and Lorentz invariance
\begin{eqnarray} 
C + C^\prime &: &\quad K, K^*_\perp, \ldots  \nonumber \\
C - C^\prime  & :& \quad K_0(1430), K^*_{0, \parallel}, \ldots
\end{eqnarray}
In addition, the $K^*_\perp$ amplitude is subleading at both high and low $q^2$ windows.
Here, $C$ and $C^\prime$ refer to V-A and  V+A quark currents, respectively, and $0, \parallel, \perp$ refers to longitudinal and transverse parallel and perpendicular transversity, respectively.
It follows that\cite{Hiller:2014ula}
\begin{eqnarray}
R_K \simeq R_\eta \simeq R_{K_1(1270,1400)}, ~~~~~R_{K^*} \simeq R_\Phi \simeq R_{K_0(1430)} \, , 
\end{eqnarray}
 and  all $R_H $ are  equal if all $C^\prime$ vanish.

Which operators are responsible for the deviation  (\ref{RKLHCb})  from universality   in $R_K,R_{K^*}$? \cite{Hiller:2017bzc}
\begin{eqnarray}
 {\rm Re}[ C_9^{{\rm NP} \mu}  -
C_{10}^{{\rm NP}  \mu}  -(\mu \to e)  ]   \sim -1.1 \pm 0.3, ~~~
  {\rm Re}[ C_9^{\prime  \mu}  - C_{10}^{\prime \mu}  -(\mu \to e)  ]  \sim  0.1 \pm 0.4  \, . 
  \end{eqnarray}
  The constraint from the  $B_s \to \mu \mu $ branching ratio  $ 0 \lsim  {\rm Re}[ 
C_{10}^{{\rm NP}  \mu} -  C_{10}^{\prime \mu}  ]  \lsim 0.9$ can be simultaneously satisfied.
The measurement of $R_K$ and $R_{K^*}$  identifies the 
V-A-type operators as the dominant source behind the anomalies.  Within leptoquark explanations,
this singles out three kinds  that can account for (\ref{RKLHCb}) at tree level: the scalar triplet leptoquark  $S_3$, the vector triplet $V_3$ and the vector singlet $V_1$, whereas the scalar doublet $\tilde S_2$ is  disfavored as it induces  V+A Wilson coefficients.
Furthermore,
LHCb data  allows one to  predict \cite{Hiller:2017bzc} $R_{X_s} \simeq  0.73 \pm 0.07$, the LNU ratio for  inclusive $B \to X_s \ell \ell$ decays, which can be probed at  Belle II.

\section{Which BSM in electrons, in muons, or in both? \label{sec:evsmu}  }

The observation of $R_H <1$ suggests
 too few muons, or too many electrons, or a combination thereof.
To disentangle this lepton specific measurements are required. Presently much more data is available on $b$-decays
to  muons than on decays to electrons.
Global $b \to s$ fits to Wilson coefficients from $B \to (K,K^*)  \mu \mu, B_s \to \mu \mu $ precision studies are presently hinting at NP, too,
and can point into the same direction as $R_{K,K^*}$. Therefore,
BSM effects in electrons are presently not necessary to account for the data.
Analogous studies  in $B \to H ee$ are, however, are required  for consolidation of this possibility. Early data are already available from Belle \cite{Wehle:2016yoi}.

Two main types of explicit BSM models can naturally address LNU at the required level of $\sim 15 \%$ on the SM amplitude:
$U(1)$ extensions with gauged lepton flavor ($Z^\prime$-models) \cite{Altmannshofer:2014cfa}  and   leptoquarks \cite{Hiller:2014yaa} \cite{crivellin}, that   can be charged under a flavor symmetry and couple non-universally \cite{Varzielas:2015iva}.

Inspection of (\ref{eq:B}) shows that 
close to maximal BSM-CP violation switches off SM-NP interference. Together with $R_H <1$ this requires large NP couplings to electrons as muons would enhance $R_H$.
Such large CP phases  in the $b \to s ee$ transition can be searched for with  the angular distribution in $B \to K^* ee$, {\it  e.g.} $J_{7,8,9}$ \cite{Hiller:2014ula}.
An explanation of $R_K$ is also possible at 2$\sigma$ with (pseudo)-scalar operators, a scenario that can be  cross checked with the $B \to K ee$ angular distribution
 \cite{Bobeth:2007dw} \cite{Hiller:2014yaa}.

\section{Side effects from flavor  \label{sec:side}}

{}From a flavor perspective, LNU generically implies LFV\cite{Glashow:2014iga}. This  is obvious for leptoquarks (LQs), which couple with matrix structure $\lambda_{q\ell}$ to quarks $q$ and leptons $\ell$
of three generations each
\begin{eqnarray} \lambda_{q \ell} =
\left( \begin{array}{ccc}
{\color{red} \lambda_{q_1 e}} & {\color{red}  \lambda_{q_1 \mu}}  & \lambda_{q_1 \tau}\\
{\color{red}  \lambda_{q_2 e}} & {\color{red}  \lambda_{q_2 \mu}}  & \lambda_{q_2 \tau}\\
\lambda_{q_3 e}  &  \lambda_{q_3 \mu} & \lambda_{q_3 \tau} 
\end{array} \right) \, ,
\label{eq:gen}
\end{eqnarray}
and rows=quarks, columns=leptons. Mixing of quark and lepton flavor in one coupling is very different from the  SM-Yukawas.
The upper left sub-matrix in red indicates the couplings relevant for  Kaon and charm physics.
Explaining $R_{K,K^*}$ requires \cite{Hiller:2017bzc}
\begin{equation} \label{eq:S3}
\frac{ \lambda_{b\mu} \lambda^\ast_{s\mu}-  \lambda_{be} \lambda^\ast_{se} }{M^2} \simeq \frac{ 1.1}{  (35\,\rm{TeV})^2} \, ,  
\end{equation}
where $M$ denotes the  LQ mass.
In matrix form, where entries with an $'\ast'$ do not matter,
\begin{eqnarray}
\label{eq:M-data}
\left(
\begin{array}{ccc}
 *  &    * &  * \\
 \lambda_{q2 e}   &   \lambda_{q2 \mu}   &  *  \\
 \lambda_{q3 e}     &  \lambda_{q3 \mu}    &  *  \\
\end{array}
\right) +~\mbox{Occam's razor ($b \to s$ fit)}:
\left(
\begin{array}{ccc}
 *  &    * &  * \\
 \ast  &   \lambda_{q2 \mu}   &  *  \\
 \ast  &  \lambda_{q3 \mu}    &  *  \\
\end{array}
\right)   \, .
\end{eqnarray}
The latter pattern assumes muon couplings only which is  consistent with the global $b \to s$ fit.
Viable patterns  from   flavor models  simultaneously  explain quark and lepton masses, and CKM and PMNS mixing \cite{Varzielas:2015iva}  \cite{Hiller:2016kry}.
For instance, models based on $U(1)_{FN} \times A_4$, with $\epsilon, \delta,  c_\ell, c_\nu \lsim 0.2$, give
\begin{eqnarray}  \label{eq:M-flavor}  \left( 
\begin{array}{ccc}
\rho_d \kappa_e &  \rho_d   & \rho_d  \kappa_\tau \\
\rho \kappa_e &  \rho   & \rho \kappa_\tau  \\
\kappa_e &  1 & \kappa_\tau
\end{array} 
\right) \, ,   \left( 
\begin{array}{ccc}
0 &  c_\ell \epsilon^4  & 0\\
0    &  c_\ell  \epsilon^2  &  0 \\
0   &  c_\ell & 0
\end{array}
 \right)  \, ,  \left( \begin{array}{ccc}c_{\nu} \kappa \epsilon^{2} & c_{\ell} \epsilon^{4} + c_{\nu} \kappa \epsilon^{2} & c_{\nu} \kappa \epsilon^{2}\\ c_{\nu} \kappa & c_{\ell} \epsilon^{2} + c_{\nu} \kappa & c_{\nu} \kappa\\ c_{\ell} \delta + c_{\nu} \kappa \epsilon^{2} & c_{\ell} & c_{\ell} \delta + c_{\nu} \kappa \epsilon^{2}\end{array} \right) \, . 
 \end{eqnarray}
 LFV and  off-diagonal couplings appear generically, as well as electron couplings, or taus.
 Phenomenological constraints apply \cite{Varzielas:2015iva}  \cite{Hiller:2016kry}.
LQs which are $SU(2)_L$ triplets couple doublets to doublets, implying BSM effects in $b \to s \nu \nu$ \cite{Hiller:2014yaa} and $b \to c \ell \nu$ \cite{Hiller:2016kry}, see section \ref{sec:RD}.
%$SU(2)$: $\nu$'s      1412.7164
%
Predictions for charm decays are given in Table \ref{tab:LQ_branching_fractions} \cite{deBoer:2015boa}.
They depend on the flavor pattern. Here, i): hierarchy, ii) muons only iii) skewed, 1) no kaon bounds 2)  kaon bounds apply for $SU(2)_L$-doublet quarks $q_2=(c,s)$.

\begin{table}[!htb]
 \centering
  \caption{\scriptsize Branching fractions for  the full $q^2$-region (high $q^2$-region) for different classes of leptoquark couplings.
     Summation of neutrino flavors is understood. 
 "SM-like" denotes a branching ratio which is dominated by resonances or is of similar size  as the resonance-induced one.
 All  $c \to u e^+ e^-$ branching ratios are "SM-like" in the models considered. See text. }
 %Note that in the SM $\mathcal B(D^0 \to \mu \mu) \sim 10^{-13}$.}
   \label{tab:LQ_branching_fractions}
 \begin{tabular}{|c|c|c|c|c|c|}
 \hline
  pattern   &  $\mathcal B(D^+\to\pi^+\mu \mu)$       &  $\mathcal B (D^0\to\mu \mu)$  &  $\mathcal B (D^+\to\pi^+e \mu)$  &  $\mathcal B (D^0\to\mu e)$  &  $\mathcal B (D^+\to\pi^+\nu\bar\nu)$  \\
  \hline
  i)   &  SM-like                                    &  SM-like                                                                                   &  $\lsim2\cdot10^{-13}$                &  $\lsim7\cdot10^{-15}$            &  $\lsim3\cdot10^{-13}$  \\
  \hline
 ii.1)  &  $\lsim7\cdot10^{-8}$ ($2\cdot10^{-8}$)                                                         &  $\lsim3\cdot10^{-9}$          &  $0$                                 &  $0$                             &  $\lsim8\cdot10^{-8}$  \\
 ii.2)  &  SM-like                                                                                             &  $\lsim4\cdot10^{-13}$         &  $0$                                 &  $0$                             &  $\lsim4\cdot10^{-12}$  \\
  \hline
 iii.1)  &  SM-like                                                                                       &  SM-like                         &  $\lsim2\cdot10^{-6}$                 & {\color{red}  $\lsim4\cdot10^{-8}$     }        &  $\lsim 2 \cdot 10^{-6}$  \\
  iii.2)  &  SM-like                                    &  SM-like                          &  $\lsim8\cdot10^{-15}$                &  $\lsim2\cdot10^{-16}$            &  $\lsim9\cdot10^{-15}$  \\
  \hline
 \end{tabular}
\end{table}

%{\scriptsize {\color{red} LHCb:  arXiv:1512.00322 [hep-ex]  ${\cal{B}}(D^0 \to e^\pm \mu^\mp) < 1.3 \cdot10^{-8}$ at 90 \% CL }}

\section{Collider implications -- leptoquarks! \label{sec:LQ} }

\begin{figure}
\begin{minipage}{0.33\linewidth}
\centerline{\includegraphics[width=1.1\linewidth]{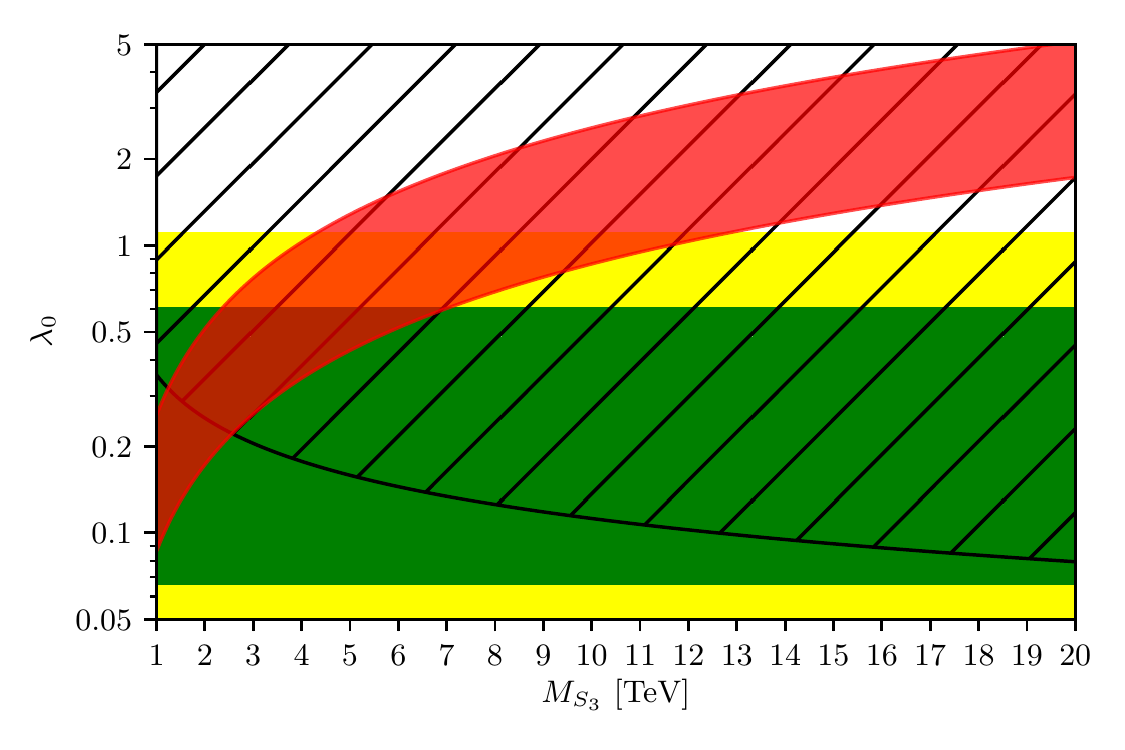}}
\end{minipage}
\hfill
\begin{minipage}{0.32\linewidth}
\centerline{\includegraphics[width=1.1\linewidth]{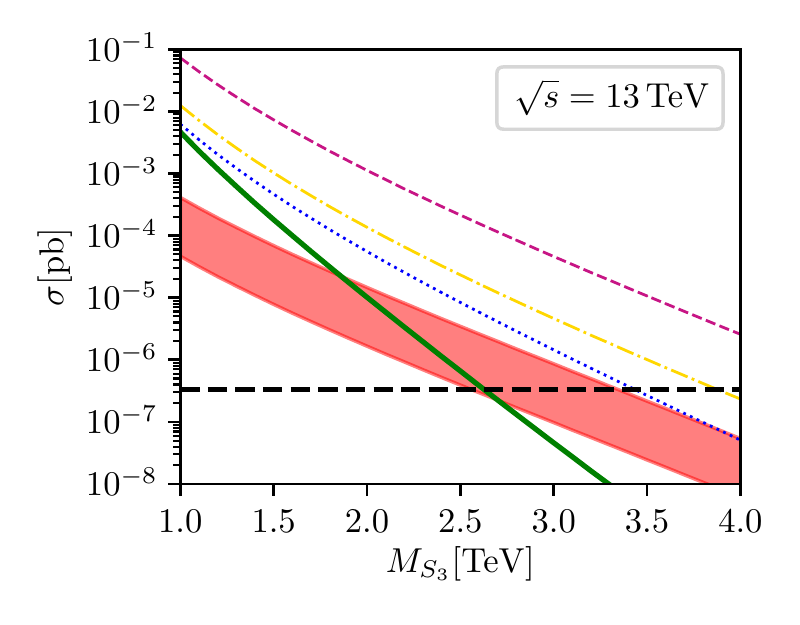}}
\end{minipage}
\hfill
\begin{minipage}{0.32\linewidth}
\centerline{\includegraphics[angle=0,width=1.1\linewidth]{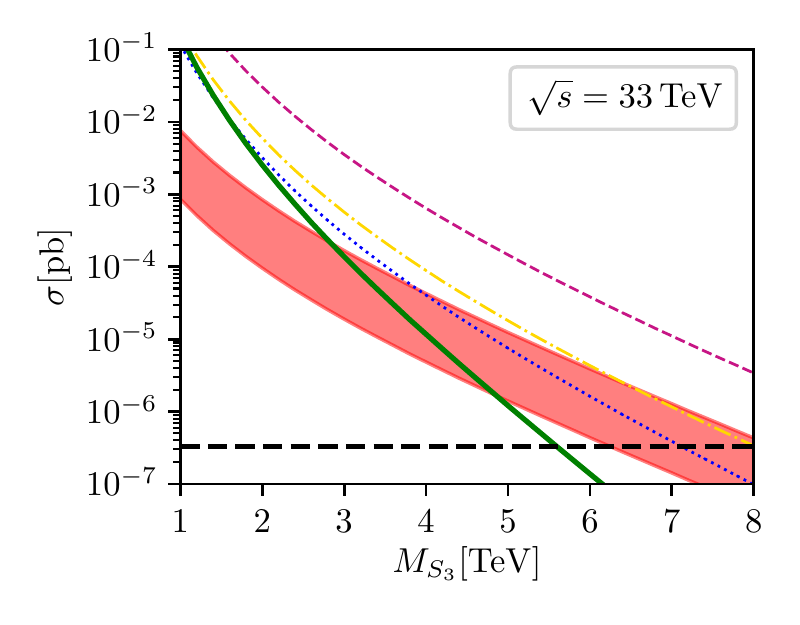}}
\end{minipage}
\caption[]{   Red bands: $R_{K,K^*}$  with flavor   (\ref{eq:blimit}).
Plot to the left shows $\lambda_{b \ell}$ vs $M$. Green vertical band gives  flavor model prediciton $\lambda_{b \ell} \sim c_\ell$ which  points to $M \lsim 7-8$ TeV.
Other plots: Single LQ production  cross section for $\sqrt{s}=13$ TeV and 33 TeV. Magenta, yellow, blue line corresponds to $\lambda_{d \mu}=1, \lambda_{s \mu}=1, \lambda_{b \mu}=1$, respectively. Black dashed line:  no-loss reach with 3 $\mbox{ab}^{-1}$.
Green curve: pair production (LO Madgraph). Figures from  \cite{Hiller:2018wbv}.  }
\label{fig:radish}
\end{figure}

Producing LQs at the LHC happens through pair production
with cross section 
$\sigma(pp \to   \phi^+ \phi^- )\propto \alpha_s^2$, recently,  {\it e.g.} \cite{Diaz:2017lit} \cite{Allanach:2017bta}  \cite{Hiller:2018wbv}.
Single LQ production 
in association with a lepton $\sigma(pp \to \phi \ell) \propto  |\lambda_{q \ell}|^2 \alpha_s$ depends on flavor, and is lesser phase space limited than pair production.
Links with $b$-anomalies and flavor are manifest  via (\ref{eq:S3})-(\ref{eq:M-flavor}).
While $b$-studies are in principle able to determine the columns, the lepton flavor structure of $\lambda_{q\ell}$,  theory input is presently required  to go on
and break the ambiguity in the product $\lambda_{b \ell} \lambda_{s \ell}^*$.
Quark hierarchies  $m_b \gg m_s \gg m_d$, when addressed with a flavor symmetry, imply hierarchies for LQs $\lambda_{s \ell } \sim (m_s/m_b) \,  \lambda_{b \ell }$.
It follows that 
third generation quark couplings dominate. Together with (\ref{eq:S3}) one obtains the range from $R_{K,K^*}$ data for  $\lambda_{b \ell}$,
\begin{eqnarray} \label{eq:blimit}
M/ 11.6\,\rm{TeV} \lsim \lambda_{b\ell} \lsim  M/ 3.9\,\rm{TeV}  \,  .
\end{eqnarray}
In figure  \ref{fig:radish} the single and pair production cross section for the scalar triplet $S_3$   is shown for $\sqrt{s}=13$ TeV and 33 TeV. One finds that
 beauty production wins  -- $bg$-fusion over $dg$- and $sg$-fusion-- also at hadronic level despite its  PDF suppression if $\lambda_{q \ell}$ follow quark mass hierarchies.
Inverted hierarchies  $\lambda_{s \ell} >  \lambda_{b \ell}$ would be surprising from a symmetry-based flavor perspective and suggest means beyond.
Looking for $pp \to \ell \ell^{(\prime)}q$ is therefore very important, yet the vanilla theory channel is $b \ell \ell^{(\prime)}$, or in pair production, $bb \ell \ell^{(\prime)}$, $\ell,\ell^\prime=e, \mu$, also LFV
$\ell \neq \ell^\prime$, and $t \ell \nu_{\ell^{(\prime)}}$.

\section{LNU in charged currents \label{sec:RD}} 

We briefly comment on the status of LNU in $b \to c \ell \nu$ decays. Input is compiled in  table \ref{tab:RD} \cite{Hiller:2016kry}.
\begin{eqnarray}
	R_{D^{(*)}}= \frac{\mathcal B(B\to D^{(*)}\tau\nu_\tau)}{\mathcal B(B\to D^{(*)}\ell\nu_\ell)}   \, , \quad \quad \hat R_{D^{(*)}} \equiv   R_{D^{(*)}}/ R_{D^{(*)}}^{\rm SM} \, , 
	\label{eq:RDdef}
\end{eqnarray}
where in the denominator of  $R_{D^{(*)}}$ $\ell=\mu$ at LHCb  and $\ell=e,\mu$ at Belle  and BaBar.
 \begin{eqnarray}  \label{eq:Rave}
	 \hat R_D^{\rm exp}=1.35 \pm 0.17 \, , \quad \quad   \hat R_{D^*}^{\rm exp}=1.23 \pm  0.07 \, ,  \quad \quad  (2016)\,   \\
	  \hat R_D^{\rm exp}=(1.35 \pm 0.17)/(1+ x) \, , \quad \quad   \hat R_{D^*}^{\rm exp}=1.18 \pm  0.07  \, ,  \quad \quad  (\mbox{NEW)}    \label{eq:RaveNEW}
 \end{eqnarray}
and $x=3.6 \%$ $(D^0)$ and $x=5.5 \%$ $(D^+)$ from  QED corrections \cite{deBoer:2018ipi}, hence $\hat R_D^{\rm exp}= 1.30 \pm 0.16$ and $1.28 \pm 0.16$, respectively \footnote{There are two caveats on the QED effects: The dependence on experimental cuts and
that the radiative corrections are not for electrons.}. See, {\it e.g.},\cite{Bernlochner:2017jka}  \cite{Jaiswal:2017rve} for  other recent SM predictions of $R_{D^{*}}$.

\begin{table}
    \centering
     \caption{Experimental results  and SM predictions for $R_D^{(*)}$,  'NEW' labels updates since 2016.  See text.
     %Table adopted and updated from \cite{Hiller:2016kry}.}
    $^\dagger$Error weighted average; we added statistical and systematical uncertainties in quadrature. }
    \begin{tabular}{|cr|c|c|}  \hline
       &     & $R_D$ & $R_{D^*}$  \\ \hline
    BaBar & \cite{Lees:2012xj} & $0.440 \pm 0.058 \pm 0.042$ & $0.332 \pm 0.024 \pm 0.018$ \\
    Belle & \cite{Huschle:2015rga}  & $0.375 \pm 0.064 \pm 0.026$ & $0.293 \pm 0.038 \pm 0.015$ \\
     Belle     & \cite{Sato:2016svk} & - & $0.302 \pm 0.030 \pm 0.011$\\
     Belle     &  \cite{Abdesselam:2016xqt}& - & $0.270 \pm 0.035 ^{+0.028}_{-0.025}$ \\
    LHCb  & \cite{Aaij:2015yra}  & - & $0.336 \pm 0.027 \pm 0.030$ \\  
         LHCb  NEW& \cite{Aaij:2017uff}  & - & $0.286 \pm 0.019 \pm 0.025 \pm 0.021$  \\ \hline
%%% average$^\dagger$ &     &  $0.406 \pm 0.050$ & $0.311 \pm 0.016$  \\ \hline
average NEW$^\dagger$ &     &  $0.406 \pm 0.050$ & $0.307 \pm 0.015$  \\ \hline
    SM &   & $0.300 \pm 0.008$  \cite{Na:2015kha}  & $0.252 \pm 0.003$ \cite{Fajfer:2012vx}   \\
    SM NEW &   & $(0.300 \pm 0.008) (1+ \%)$  \cite{deBoer:2018ipi}  & $0.260 \pm 0.008$ \cite{Bigi:2017jbd}  \\
    \hline
    \end{tabular}
        \label{tab:RD}
\end{table}
In some scenarios, such as LQs $S_3,V_3$ and $V_1$ BSM effects in $R_{K,K^*}$ imply BSM effects in $R_{D,D^*}$, however, due to the large  SM contribution in the tree level decays, at a
reduced level. Flavor models predict effects up to few percent and  around 10 percent in $R_{D^*}$ and $R_D$ \cite{Hiller:2016kry}, respectively, below the present 1 $\sigma$ ranges,
(\ref{eq:Rave})-(\ref{eq:RaveNEW}).

\section{Summary}

Current data on $R_K, R_{K^*},R_D, R_{D^*}$ in semileptonic $B$-meson decays hint at violation of lepton-universality, and therefore the breakdown of the SM.
The April 2017 release of $R_{K^*}$ by LHCb has strengthened the previous hints and allowed to pin down the Dirac structure of the underlying physics to be  predominantly of V-A-type.
Future data -- LNU updates and  other observables $R_{\Phi}, R_{Xs}...,B\to K^{(*)} ee$ -- from LHCb and in the nearer future from Belle II  are eagerly awaited. 

What makes these LNU-anomalies -- iff true -- so important? They are theoretically clean and intimately linked to flavor:  
they can give new insights towards the origin of flavor
and  structure by probing models of flavor. Correspondingly, one  should look for imprints in other sectors: $D$, $K$ physics, LFV, including $\mu-e$ conversion and lepton decays.

In addition, new BSM model buildung  has been triggered that deserves attention in direct searches at ATLAS and CMS and future colliders.
Leptoquarks are flavorful and can be in reach of the LHC, where they can provide complementary information to rare decays, on the  couplings $\lambda_{s \ell}, \lambda_{b \ell}$ and masses
$ M$ separately  vs  their product (\ref{eq:S3}).
Model-independent upper limits on $M$ are at the few ${\cal{O}}(10)$ TeV level, $40, 45$ and 20 TeV  for $S_3,V_1$ and $V_3$, respectively \cite{Hiller:2017bzc}.
% by $B_s$-mixing $\propto (\lambda_{b \ell} \lambda_{s \ell})^2/M^2$ at  $\sim40$ TeV.
The bulk of the  parameter space lies outside of the LHC  \cite{Allanach:2017bta} \cite{Hiller:2018wbv}.

\vspace{-0.1cm}

\section*{Acknowledgments}

GH is happy to thank the organizers for the  opportunity to speak at this wonderful conference,
and her collaborators for great collaboration. This work is supported in part by the {\it Bundesministerium f\"ur Bildung und Forschung} (BMBF).


\section*{References}

\begin{thebibliography}{99}

%\cite{Hiller:2003js}
\bibitem{Hiller:2003js} 
  G.~Hiller and F.~Kr\"uger,
  %``More model-independent analysis of $b \to s$ processes,''
  Phys.\ Rev.\ D {\bf 69}, 074020 (2004)
  doi:10.1103/PhysRevD.69.074020
  [hep-ph/0310219].
  %%CITATION = doi:10.1103/PhysRevD.69.074020;%%
  
  
  %\cite{Aaij:2014ora}
\bibitem{Aaij:2014ora}
  R.~Aaij {\it et al.} [LHCb Collaboration],
  %``Test of lepton universality using $B^{+}\rightarrow K^{+}\ell^{+}\ell^{-}$ decays,''
  Phys.\ Rev.\ Lett.\  {\bf 113} (2014) 151601
  %doi:10.1103/PhysRevLett.113.151601
  [arXiv:1406.6482 [hep-ex]].
  %%CITATION = doi:10.1103/PhysRevLett.113.151601;%%
  
   %\cite{Aaij:2017vbb}
\bibitem{Aaij:2017vbb} 
  R.~Aaij {\it et al.} [LHCb Collaboration],
  %``Test of lepton universality with $B^{0} \rightarrow K^{*0}\ell^{+}\ell^{-}$ decays,''
  JHEP {\bf 1708}, 055 (2017)
  %doi:10.1007/JHEP08(2017)055
  [arXiv:1705.05802 [hep-ex]].
  %%CITATION = doi:10.1007/JHEP08(2017)055;%%
  
  \bibitem{albrecht} J.~Albrecht, {\it these proceedings}.
  
  %\cite{Huber:2005ig}
\bibitem{Huber:2005ig} 
  T.~Huber, E.~Lunghi, M.~Misiak and D.~Wyler,
  %``Electromagnetic logarithms in $\bar B \to  X_s l^+ l^-$,''
  Nucl.\ Phys.\ B {\bf 740}, 105 (2006)
  doi:10.1016/j.nuclphysb.2006.01.037
  [hep-ph/0512066].
  %%CITATION = doi:10.1016/j.nuclphysb.2006.01.037;%%
  
  %\cite{Bobeth:2007dw}
\bibitem{Bobeth:2007dw} 
  C.~Bobeth, G.~Hiller and G.~Piranishvili,
  %``Angular distributions of $\bar{B} \to \bar{K} \ell^+\ell^-$ decays,''
  JHEP {\bf 0712}, 040 (2007)
  doi:10.1088/1126-6708/2007/12/040
  [arXiv:0709.4174 [hep-ph]].
  %%CITATION = doi:10.1088/1126-6708/2007/12/040;%%
  
  
%  \cite{Bordone:2016gaq}
\bibitem{Bordone:2016gaq} 
  M.~Bordone, G.~Isidori and A.~Pattori,
  %``On the Standard Model predictions for $R_K$ and $R_{K^*}$,''
  Eur.\ Phys.\ J.\ C {\bf 76}, no. 8, 440 (2016)
  doi:10.1140/epjc/s10052-016-4274-7
  [arXiv:1605.07633 [hep-ph]].
  %%CITATION = doi:10.1140/epjc/s10052-016-4274-7;%%
  
  
  %\cite{Hiller:2014yaa}
\bibitem{Hiller:2014yaa} 
  G.~Hiller and M.~Schmaltz,
  %``$R_K$ and future $b \to s \ell \ell$ physics beyond the standard model opportunities,''
  Phys.\ Rev.\ D {\bf 90}, 054014 (2014)
  doi:10.1103/PhysRevD.90.054014
  [arXiv:1408.1627 [hep-ph]].
  %%CITATION = doi:10.1103/PhysRevD.90.054014;%%
  
  %\cite{Hiller:2014ula}
\bibitem{Hiller:2014ula} 
  G.~Hiller and M.~Schmaltz,
  %``Diagnosing lepton-nonuniversality in $b \to s \ell \ell$,''
  JHEP {\bf 1502}, 055 (2015)
  doi:10.1007/JHEP02(2015)055
  [arXiv:1411.4773 [hep-ph]].
  %%CITATION = doi:10.1007/JHEP02(2015)055;%%



\bibitem{fleischer} R.~Fleischer, {\it these proceedings}.

\bibitem{crocombe} A.~Crocombe, {\it these proceedings}.

  %\cite{Glashow:2014iga}
\bibitem{Glashow:2014iga} 
  S.~L.~Glashow, D.~Guadagnoli and K.~Lane,
  %``Lepton Flavor Violation in $B$ Decays?,''
  Phys.\ Rev.\ Lett.\  {\bf 114}, 091801 (2015)
  doi:10.1103/PhysRevLett.114.091801
  [arXiv:1411.0565 [hep-ph]].
  %%CITATION = doi:10.1103/PhysRevLett.114.091801;%%
  
  %\cite{Varzielas:2015iva}
\bibitem{Varzielas:2015iva} 
  I.~de Medeiros Varzielas and G.~Hiller,
  %``Clues for flavor from rare lepton and quark decays,''
  JHEP {\bf 1506}, 072 (2015)
  doi:10.1007/JHEP06(2015)072
  [arXiv:1503.01084 [hep-ph]].
  %%CITATION = doi:10.1007/JHEP06(2015)072;%%

\bibitem{iyer} A.~Iyer, {\it these proceedings}.

\bibitem{GD} G.~d'Ambrosio, {\it these proceedings}.

\bibitem{you} T.~You, {\it these proceedings}.

\bibitem{morse} D.M.~Morse, {\it these proceedings}.

\bibitem{crivellin} A.~Crivellin, {\it these proceedings}.



%\cite{Hiller:2017bzc}
\bibitem{Hiller:2017bzc} 
  G.~Hiller and I.~Ni\v{s}and\v{z}i\'c,
  %``$R_K$ and $R_{K^{\ast}}$ beyond the standard model,''
  Phys.\ Rev.\ D {\bf 96}, no. 3, 035003 (2017)
  doi:10.1103/PhysRevD.96.035003
  [arXiv:1704.05444 [hep-ph]].
  %%CITATION = doi:10.1103/PhysRevD.96.035003;%%
  
  %\cite{Wehle:2016yoi}
\bibitem{Wehle:2016yoi} 
  S.~Wehle {\it et al.} [Belle Collaboration],
  %``Lepton-Flavor-Dependent Angular Analysis of $B\to K^\ast \ell^+\ell^-$,''
  Phys.\ Rev.\ Lett.\  {\bf 118}, no. 11, 111801 (2017)
  doi:10.1103/PhysRevLett.118.111801
  [arXiv:1612.05014 [hep-ex]].
  %%CITATION = doi:10.1103/PhysRevLett.118.111801;%%
  
  %\cite{Altmannshofer:2014cfa}
\bibitem{Altmannshofer:2014cfa} 
  W.~Altmannshofer, S.~Gori, M.~Pospelov and I.~Yavin,
  %``Quark flavor transitions in $L_\mu-L_\tau$ models,''
  Phys.\ Rev.\ D {\bf 89}, 095033 (2014)
  doi:10.1103/PhysRevD.89.095033
  [arXiv:1403.1269 [hep-ph]].
  %%CITATION = doi:10.1103/PhysRevD.89.095033;%%


%\cite{Hiller:2016kry}
\bibitem{Hiller:2016kry} 
  G.~Hiller, D.~Loose and K.~Sch\"onwald,
  %``Leptoquark Flavor Patterns & B Decay Anomalies,''
  JHEP {\bf 1612}, 027 (2016)
  doi:10.1007/JHEP12(2016)027
  [arXiv:1609.08895 [hep-ph]].
  %%CITATION = doi:10.1007/JHEP12(2016)027;%%
  
  
    

  
  %\cite{deBoer:2015boa}
\bibitem{deBoer:2015boa} 
  S.~de Boer and G.~Hiller,
  %``Flavor and new physics opportunities with rare charm decays into leptons,''
  Phys.\ Rev.\ D {\bf 93}, no. 7, 074001 (2016)
  doi:10.1103/PhysRevD.93.074001
  [arXiv:1510.00311 [hep-ph]].
  %%CITATION = doi:10.1103/PhysRevD.93.074001;%%
  
   %\cite{Diaz:2017lit}
\bibitem{Diaz:2017lit} 
  B.~Diaz, M.~Schmaltz and Y.~M.~Zhong,
  %``The leptoquark HunterÕs guide: Pair production,''
  JHEP {\bf 1710}, 097 (2017)
  doi:10.1007/JHEP10(2017)097
  [arXiv:1706.05033 [hep-ph]].
  %%CITATION = doi:10.1007/JHEP10(2017)097;%%
  
  %\cite{Allanach:2017bta}
\bibitem{Allanach:2017bta} 
  B.~C.~Allanach, B.~Gripaios and T.~You,
  %``The case for future hadron colliders from $B \to K^{(*)} \mu^+ \mu^-$  decays,''
  JHEP {\bf 1803}, 021 (2018)
  doi:10.1007/JHEP03(2018)021
  [arXiv:1710.06363 [hep-ph]].
  %%CITATION = doi:10.1007/JHEP03(2018)021;%%

  
  %\cite{Hiller:2018wbv}
\bibitem{Hiller:2018wbv} 
  G.~Hiller, D.~Loose and I.~Ni\v{s}and\v{z}i\'c,
  %``Flavorful leptoquarks at hadron colliders,''
  Phys. Rev. D 97, 075004 (2018)
  doi:10.1103/PhysRevD.97.075004
 [ arXiv:1801.09399 [hep-ph]].
  %%CITATION = ARXIV:1801.09399;%%
  
  %\cite{Abdesselam:2016xqt}
\bibitem{Abdesselam:2016xqt} 
%\cite{Hirose:2016wfn}
%%\bibitem{Hirose:2016wfn} 
  S.~Hirose {\it et al.} [The Belle Collaboration],
  %``Measurement of the $\tau$ lepton polarization and $R(D^*)$ in the decay $\bar{B} \to D^* \tau^- \bar{\nu}_\tau$,''
  arXiv:1612.00529 [hep-ex].
  %%CITATION = ARXIV:1612.00529;%%



%\cite{Lees:2012xj}
\bibitem{Lees:2012xj} 
  J.~P.~Lees {\it et al.} [BaBar Collaboration],
  %``Evidence for an excess of $\bar{B} \to D^{(*)} \tau^-\bar{\nu}_\tau$ decays,''
  Phys.\ Rev.\ Lett.\  {\bf 109}, 101802 (2012)
  doi:10.1103/PhysRevLett.109.101802
  [arXiv:1205.5442 [hep-ex]].
  %%CITATION = doi:10.1103/PhysRevLett.109.101802;%%
  %322 citations counted in INSPIRE as of 26 Sep 2016



%\cite{Huschle:2015rga}
\bibitem{Huschle:2015rga} 
  M.~Huschle {\it et al.} [Belle Collaboration],
  %``Measurement of the branching ratio of $\bar{B} \to D^{(\ast)} \tau^- \bar{\nu}_\tau$ relative to $\bar{B} \to D^{(\ast)} \ell^- \bar{\nu}_\ell$ decays with hadronic tagging at Belle,''
  Phys.\ Rev.\ D {\bf 92}, no. 7, 072014 (2015)
  doi:10.1103/PhysRevD.92.072014
  [arXiv:1507.03233 [hep-ex]].
  %%CITATION = doi:10.1103/PhysRevD.92.072014;%%
  %98 citations counted in INSPIRE as of 26 Sep 2016



%\cite{Sato:2016svk}
\bibitem{Sato:2016svk} 
  Y.~Sato {\it et al.} [Belle Collaboration],
  %``Measurement of the branching ratio of $\bar{B}^0 \rightarrow D^{*+} \tau^- \bar{\nu}_{\tau}$ relative to $\bar{B}^0 \rightarrow D^{*+} \ell^- \bar{\nu}_{\ell}$ decays with a semileptonic tagging method,''
  arXiv:1607.07923 [hep-ex].
  %%CITATION = ARXIV:1607.07923;%%
  %2 citations counted in INSPIRE as of 26 Sep 2016



%\cite{Aaij:2015yra}
\bibitem{Aaij:2015yra} 
  R.~Aaij {\it et al.} [LHCb Collaboration],
  %``Measurement of the ratio of branching fractions $\mathcal{B}(\bar{B}^0 \to D^{*+}\tau^{-}\bar{\nu}_{\tau})/\mathcal{B}(\bar{B}^0 \to D^{*+}\mu^{-}\bar{\nu}_{\mu})$,''
  Phys.\ Rev.\ Lett.\  {\bf 115}, no. 11, 111803 (2015)
  Addendum: [Phys.\ Rev.\ Lett.\  {\bf 115}, no. 15, 159901 (2015)]
  doi:10.1103/PhysRevLett.115.159901, 10.1103/PhysRevLett.115.111803
  [arXiv:1506.08614 [hep-ex]].
  %%CITATION = doi:10.1103/PhysRevLett.115.159901, 10.1103/PhysRevLett.115.111803;%%
  %115 citations counted in INSPIRE as of 26 Sep 2016



%\cite{Fajfer:2012vx}
\bibitem{Fajfer:2012vx} 
  S.~Fajfer, J.~F.~Kamenik and I.~Ni\v{s}and\v{z}i\'c,
  %``On the $B \to D^* \tau \bar \nu_{\tau}$ Sensitivity to New Physics,''
  Phys.\ Rev.\ D {\bf 85}, 094025 (2012)
  doi:10.1103/PhysRevD.85.094025
  [arXiv:1203.2654 [hep-ph]].
  %%CITATION = doi:10.1103/PhysRevD.85.094025;%%
  %180 citations counted in INSPIRE as of 26 Sep 2016

%\cite{Na:2015kha}
\bibitem{Na:2015kha} 
  H.~Na {\it et al.} [HPQCD Collaboration],
  %``$B \rightarrow D l \nu$ form factors at nonzero recoil and extraction of $|V_{cb}|$,''
  Phys.\ Rev.\ D {\bf 92}, no. 5, 054510 (2015)
  Erratum: [Phys.\ Rev.\ D {\bf 93}, no. 11, 119906 (2016)]
  doi:10.1103/PhysRevD.93.119906, 10.1103/PhysRevD.92.054510
  [arXiv:1505.03925 [hep-lat]].
  %%CITATION = doi:10.1103/PhysRevD.93.119906, 10.1103/PhysRevD.92.054510;%%
  %32 citations counted in INSPIRE as of 26 Sep 2016



%\cite{Aaij:2017uff}
\bibitem{Aaij:2017uff} 
  R.~Aaij {\it et al.} [LHCb Collaboration],
  %``Measurement of the ratio of the $B^0 \to D^{*-} \tau^+ \nu_{\tau}$ and $B^0 \to D^{*-} \mu^+ \nu_{\mu}$ branching fractions using three-prong $\tau$-lepton decays,''
  arXiv:1708.08856 [hep-ex].
  %%CITATION = ARXIV:1708.08856;%%
  
  %\cite{deBoer:2018ipi}
\bibitem{deBoer:2018ipi} 
  S.~de Boer, T.~Kitahara and I.~Ni\v{s}and\v{z}i\'c,
  %``Soft-photon corrections to $\bar{B} \to D \tau^{-} \bar{\nu}_{\tau}$ relative to $\bar{B} \to D \mu^{-} \bar{\nu}_{\mu}$,''
  arXiv:1803.05881 [hep-ph].
  %%CITATION = ARXIV:1803.05881;%%

  
  %\cite{Bigi:2017jbd}
\bibitem{Bigi:2017jbd} 
  D.~Bigi, P.~Gambino and S.~Schacht,
  %``$R(D^*)$, $|V_{cb}|$, and the Heavy Quark Symmetry relations between form factors,''
  JHEP {\bf 1711}, 061 (2017)
  doi:10.1007/JHEP11(2017)061
  [arXiv:1707.09509 [hep-ph]].
  %%CITATION = doi:10.1007/JHEP11(2017)061;%%
  
  %\cite{Bernlochner:2017jka}
\bibitem{Bernlochner:2017jka} 
  F.~U.~Bernlochner, Z.~Ligeti, M.~Papucci and D.J.~Robinson,
  %``Combined analysis of semileptonic $B$ decays to $D$ and $D^*$: $R(D^{(*)})$, $|V_{cb}|$, and new physics,''
  Phys.~Rev.~D {\bf95}, no.11, 115008 (2017)~Erratum: [Phys.~Rev.~D {\bf97}, no. 5, 059902 (2018)]
  doi:10.1103/PhysRev\\D.95.115008,10.1103/PhysRevD.97.059902~[arXiv:1703.05330 [hep-ph]].
  %%CITATION = doi:10.1103/PhysRevD.95.115008, 10.1103/PhysRevD.97.059902;%%
  
  %\cite{Jaiswal:2017rve}
\bibitem{Jaiswal:2017rve}
 S.~Jaiswal, S.~Nandi and S.~K.~Patra,
 %``Extraction of $|V_{cb}|$ from $B\to D^{(*)}\ell\nu_\ell$ and the
Standard Model predictions of $R(D^{(*)})$,''
 JHEP {\bf 1712}, 060 (2017)
 doi:10.1007/JHEP12(2017)060
 [arXiv:1707.09977 [hep-ph]].
 %%CITATION = doi:10.1007/JHEP12(2017)060;%%
  
 
\end{thebibliography}
\end{document}